\documentclass[conference]{IEEEtran}
\IEEEoverridecommandlockouts
\usepackage{cite}
\usepackage{amsmath,amssymb,amsfonts}
\usepackage{graphicx}
\usepackage{textcomp}
\usepackage{xcolor}

\usepackage{subfig}
\usepackage{algorithm}
\usepackage{algpseudocode}
\usepackage{multirow}
\usepackage{makecell}
\usepackage{url}
\usepackage{newtxtext,newtxmath}
\usepackage{diagbox}

\def\BibTeX{{\rm B\kern-.05em{\sc i\kern-.025em b}\kern-.08em
    T\kern-.1667em\lower.7ex\hbox{E}\kern-.125emX}}
\begin{document}

\title{MixServe: An Automatic Distributed Serving System for MoE Models with Hybrid Parallelism Based on Fused Communication Algorithm}

\author{\IEEEauthorblockN{Bowen Zhou\textsuperscript{1}, Jinrui Jia\textsuperscript{1}, Wenhao He\textsuperscript{1}, Yong Zhang\textsuperscript{2}, Fang Dong\textsuperscript{1}\textsuperscript{*}}
\IEEEauthorblockA{\textsuperscript{1}School of Computer Science and Engineering, Southeast University, Nanjing, China \\
\textsuperscript{2}Department of Computer Science and Engineering, The Chinese University of Hong Kong, Hong Kong, China \\
Emails: \{bwzhou, jrjia, wenhao\_he\}@seu.edu.cn, zhangyong@link.cuhk.edu.hk, fdong@seu.edu.cn \\
\thanks{\textsuperscript{*}Corresponding author.}}
}

\maketitle
\bstctlcite{IEEEexample:BSTcontrol}

\begin{abstract}
  The Mixture of Experts (MoE) models are emerging as the latest paradigm for Large Language Models (LLMs). However, due to memory constraints, MoE models with billions or even trillions of parameters can only be deployed in multi-GPU or even multi-node \& multi-GPU based serving systems. Thus, communication has became a major bottleneck in distributed serving systems, especially inter-node communication. Contemporary distributed MoE models are primarily implemented using all-reduce (AR) based tensor parallelism (TP) and all-to-all (A2A) based expert parallelism (EP). However, TP generally exhibits low inter-node efficiency and is thus confined to high-speed intra-node bandwidth. In contrast, EP tends to suffer from load imbalance, especially when the parallel degree is high.

  In this work, we introduce \textbf{MixServe}, a novel automatic distributed serving system for efficient deployment of MoE models by a novel TP-EP hybrid parallelism based on fused AR-A2A communication algorithm. MixServe begins by evaluating the communication overhead associated with various parallel strategies, taking into account the model hyperparameters and the configurations of network and hardware resources, and then automatically selects the most efficient parallel strategy. Then, we propose the TP-EP hybrid parallelism based on fused AR-A2A communication algorithm that overlaps intra-node AR communication and inter-node A2A communication. Extensive experiments on DeepSeek-R1 and Qwen3 models demonstrate that MixServe achieves superior inference performance, with 1.08 $\times$ $\sim$ 3.80 $\times$ acceleration in time to first token (TTFT), 1.03 $\times$ $\sim$ 1.66 $\times$ acceleration in inter-token latency (ITL), and 5.2\% $\sim$ 50.3\% throughput improvement compared to existing approaches.
\end{abstract}

% Keywords

% add the paper content here
\section{Introduction}

Mixture of Experts (MoE) models (\textit{e.g.}, DeepSeek-V3 \cite{DeepSeek-V3}, DeepSeek-R1 \cite{DeepSeek-R1}, Qwen3 \cite{Qwen3}) are emerging as the latest paradigm for Large Language Models (LLMs). Benifiting from the sparse activation mechanism, MoE models can scale up the model capacity to tens of billions, hundreds of billions, or even trillions of parameters, so that they can achieve competitive performance while maintaining inference-time computational efficiency. By leveraging a routing mechanism that activates only a subset of model parameters (experts) for each input token, MoE models achieve competitive accuracy while maintaining inference-time computational efficiency. This sparse activation strategy enables MoE models to reach parameter counts in the tens of billions, hundreds of billions, or even trillions while keeping the activated parameters manageable.

However, the increasing scale and complexity of MoE models pose significant challenges for achieving efficient inference in real-world deployment scenarios. Due to their massive parameter counts, these models can only be deployed in multi-GPU or even multi-node \& multi-GPU distributed environments, where communication becomes a critical performance bottleneck, especially inter-node communication. Although intra-node high-speed interconnect protocols (\textit{e.g.}, NVLink) offer substantial bandwidth advantages, inter-node connections typically rely on protocols like InfiniBand or RoCE\footnote{RDMA over Converged Ethernet} that provide bandwidth significantly lower than intra-node interconnects. The disparity in bandwidth necessitates that high-frequency or extensive inter-node communication must be minimized during the design of communication algorithms, as this has emerged as a significant bottleneck in distributed MoE model service systems.

Contemporary distributed MoE implementations primarily rely on two main parallelism strategies for handling the computational and communication demands: tensor parallelism (TP) using all-reduce (AR) operators, and expert parallelism (EP) using all-to-all (A2A) operators. TP distributes model parameters across multiple devices and requires frequent synchronization through AR operators, which typically performs well within a single node due to high-speed intra-node interconnects but scales poorly across nodes due to limited inter-node bandwidth. In contrast, EP distributes different experts across multiple devices and uses A2A operators to route tokens to their assigned experts, offering better inter-node scaling potential but often suffering from load imbalance and communication inefficiencies, especially when the parallel degree is high.

Building on these observations, current deployment strategies for MoE models typically adopt a hybrid approach: utilizing TP for the Attention blocks and EP for the MoE blocks, utilizing pipeline parallelism (PP) to shard the Decoders \cite{liu2025moeparallelfoldingheterogeneous}. This division is motivated by the fact that Attention blocks generally contain fewer parameters than MoE blocks, making them suitable for TP with manageable communication overhead. Meanwhile, data parallelism (DP) can be effectively implemented within the Attention blocks to improve overall system throughput. For instance, in the DeepSeek-V3 technical report \cite{DeepSeek-V3}, the authors advocate that MoE blocks should fully adopt EP to ensure that each expert can process sufficiently large batch sizes, thereby maximizing computational efficiency. However, existing parallel strategies suffer from the following critical limitations:

\textbf{Lack of Systematic Theoretical Analysis}. Existing parallel strategies are primarily based on empirical intuition and practical experience, lacking comprehensive theoretical analysis and systematic validation. These approaches fail to adequately consider the complex interplay between model hyperparameters, network topology, and hardware resource configurations.

\textbf{Ineffective Exploitation of Communication Bandwidth Hierarchies}. Most critically, current approaches do not effectively exploit the significant bandwidth disparities that exist between intra-node and inter-node communication channels. Current communication operators treat all communication uniformly, missing opportunities to optimize performance by leveraging the heterogeneous nature of distributed computing environments.

To address these fundamental communication and scalability challenges, we introduce MixServe, a novel automatic distributed serving system designed specifically for efficient deployment of MoE models. MixServe employs a systematic approach that begins by comprehensively evaluating the communication overhead associated with various parallel strategies, taking into account not only the hyperparameters of the MoE model but also the detailed configuration of network topology and hardware resources. Based on this analysis, MixServe automatically selects the most efficient parallel strategy for each specific deployment scenario.

The core innovation of MixServe lies in its TP-EP hybrid parallelism, implemented through a sophisticated fused AR-A2A communication algorithm. This algorithm strategically overlaps intra-node AR communication with inter-node A2A communication, effectively exploiting the bandwidth hierarchy present in modern distributed systems. By carefully orchestrating these overlapping communication patterns, MixServe minimizes overall communication latency while maintaining computational correctness. Our comprehensive evaluation on large-scale MoE models, including DeepSeek-R1 and Qwen3, demonstrates that MixServe consistently achieves superior inference performance compared to state-of-the-art approaches. Specifically, MixServe delivers 1.08 $\times$ $\sim$ 3.80 $\times$ acceleration in time to first token (TTFT), 1.03 $\times$ $\sim$ 1.66 $\times$ acceleration in inter-token latency (ITL), and 5.2\% $\sim$ 50.3\% improvement in overall throughput. By bridging the gap between system-level efficiency and model scalability, MixServe enables practical, high-performance deployment of MoE models in real-world distributed serving environments. Our contributions can be summarized as follows:

\begin{itemize}
  \item \textbf{Automatic Serving System}: We present MixServe, a novel automatic distributed serving system that systematically evaluates communication overhead and automatically selects optimal parallel strategies based on model hyperparameters and network configurations, replacing empirical intuition with rigorous analysis.

  \item \textbf{Theoretical Communication Analysis}: We conduct comprehensive theoretical analysis of communication overhead for distributed MoE serving, deriving rigorous models for TP, PP, EP, and DP strategies that enable informed strategy selection based on hardware characteristics and network bandwidth hierarchies.

  \item \textbf{Fused AR-A2A Communication Algorithm}: We propose a novel TP-EP hybrid parallelism with fused AR-A2A communication algorithm that efficiently overlaps intra-node AR communication with inter-node A2A communication, significantly reducing overall communication latency.

  \item \textbf{Performance Evaluation on Mainstream MoEs}: We demonstrate substantial performance improvements on DeepSeek-R1 and Qwen3 models: $1.08 \times \sim 3.80 \times$ TTFT acceleration, $1.03 \times \sim 1.66 \times$ ITL acceleration, and 5.2\% $\sim$ 50.3\% throughput improvement compared to existing approaches.
\end{itemize}

\section{Background \& Motivation}

\subsection{Hybrid Parallelism for MoE Models}

\begin{figure}[t]
  \centering
  \subfloat[TP]{
    \includegraphics[height=0.15\textheight]{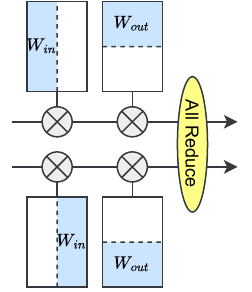}
    \label{fig:TP}
  }
  \subfloat[AR]{
    \includegraphics[height=0.15\textheight]{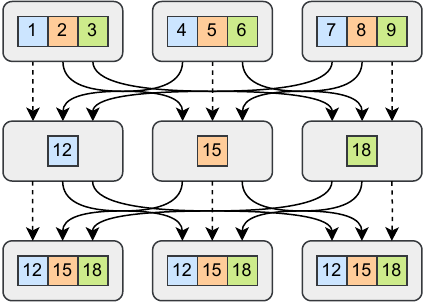}
    \label{fig:AR}
  }
  \caption{Tensor Parallelism (TP) and All Reduce (AR) operators.}
\end{figure}

\begin{figure}[t]
  \centering
  \subfloat[EP]{
    \includegraphics[height=0.09\textheight]{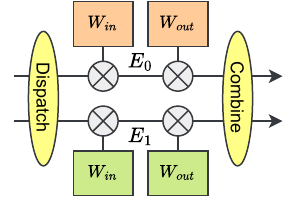}
    \label{fig:EP}
  }
  \subfloat[A2A]{
    \includegraphics[height=0.09\textheight]{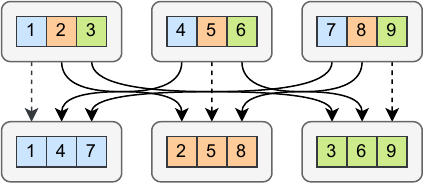}
    \label{fig:A2A}
  }
  \caption{Expert Parallelism (EP) and All To All (A2A) operators.}
\end{figure}

The core component of the MoE model primarily utilizes a hybrid TP and EP parallelism strategy. TP refers to the partitioning of model parameters across multiple nodes in a distributed system, initially proposed in Megatron-LM \cite{Megatron-LM}. Each node holds a replica of a subset of the model parameters, and they are updated independently using AR. AR aggregates gradients from all nodes and updates the parameters in a synchronous manner. This approach enables efficient training of large-scale models by distributing the workload across multiple nodes. Fig.~\ref{fig:TP} illustrates the execution process of TP, which sums up the hidden states after $W^{\mathrm{in}}$ and $W^{\mathrm{out}}$ projection. Fig.~\ref{fig:AR} shows the execution process of AR. It is usually divided into two operators: (1) Reduce Scatter (RS), (2) All Gather (AG), because it has lower communication overhead compared to direct implementations. In a Decoder-structured LLM, TP result synchronization is performed once during both the forward and backward propagation phases of the Attention and Feed-Forward Network (FFN) blocks. As a result, TP is generally unsuitable for inter-node implementation and is typically executed within a single node equipped with high-speed interconnect protocols.

EP refers to the partitioning of model computation across multiple experts in a MoE model, first explicitly proposed in Switch Transformer \cite{Switch-Transformer}. Each expert holds a replica of a subset of the model computation, and they are updated independently using A2A operators. A2A exchanges gradients between nodes and updates the parameters in an asynchronous manner. This approach enables efficient inference of large-scale MoE models by distributing the computation across multiple experts. In the DeepSeek-V3 technical report \cite{DeepSeek-V3}, researchers assert that the MoE block should fully adopt EP to ensure that each expert can process sufficiently large batch sizes. Fig.~\ref{fig:EP} illustrates the execution process of EP, where each token is sent to the device hosting the corresponding expert (\textit{i.e.} Dispatch). Upon completing inference through the MLP layer, the results are returned along the original route (\textit{i.e.} Combine). Fig.~\ref{fig:A2A} shows the execution process of A2A. A2A communication can be implemented through various algorithms, among which Ring and Pairwise are commonly used. Both require $N-1$ rounds ($N$ denotes the total number of participating devices) to complete the transmission and reception of all data.

\subsection{Challenges of Inter-node Communication}

\begin{figure}[t]
  \centering
  \includegraphics[width=0.45\textwidth]{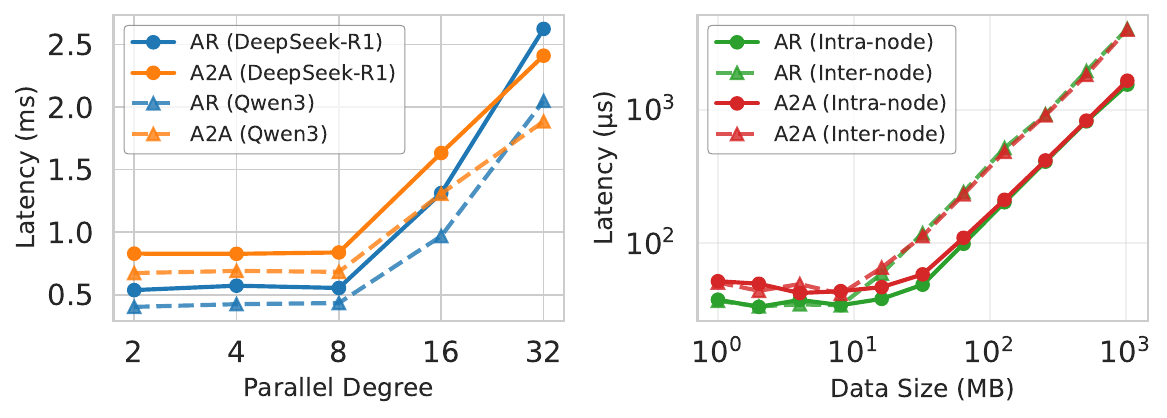}
  \caption{Communication overhead of AR and A2A operators. The left subfigure shows the communication overhead for DeepSeek-R1 \cite{DeepSeek-R1} and Qwen3 \cite{Qwen3} models with different parallel degrees, while the right subfigure presents the results for intra-node and inter-node communication with different data sizes.}
  \label{fig:comm_ar_a2a}
\end{figure}

\begin{figure}[t]
  \centering
  \includegraphics[width=0.45\textwidth]{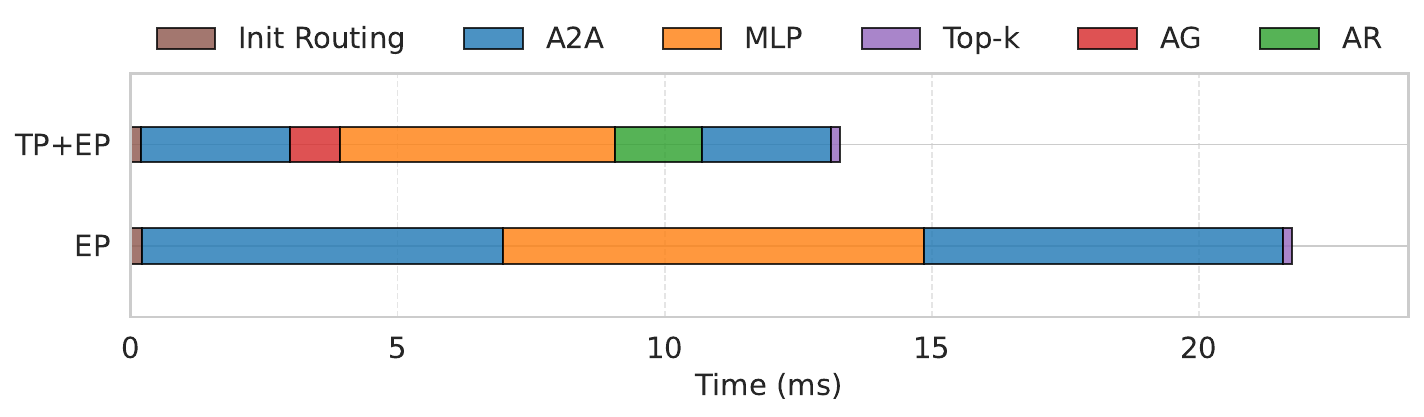}
  \caption{Gantt chart of the comparison between EP and TP+EP in a single MoE block, calculated from profiling data of DeepSeek-R1 in a 4-node cluster with 8 NPUs each.}
  \label{fig:moe_ep_vs_tp_ep_gantt}
\end{figure}

The deployment of large-scale MoE models for inference introduces significant system-level challenges. As the model parameters and experts increases, multi-GPU or even multi-node \& multi-GPU deployment is required to meet the computational demands. However, the communication overhead in distributed environments quickly becomes the dominant bottleneck. Fig.~\ref{fig:comm_ar_a2a} illustrates the communication overhead of AR and A2A operators, which is obtained in a cluster of 4 Atlas 800T A2 servers with 8 NPUs each. Initially, we collected the profiling data of the DeepSeek-R1 \cite{DeepSeek-R1} and Qwen3-235B-A22B \cite{Qwen3} models under two configurations of the MoE block, specifically using TP and EP. Subsequently, we extracted the latency measurements for the two operators, AR and A2A, as shown in the left subfigure of Fig.~\ref{fig:comm_ar_a2a}. We denote $d$ as the degree of parallelism for different parallel strategies, referring to the total number of devices participating in the communication. When $d \leq 8$, it signifies that communication is conducted entirely within the node. Experimental results indicate that the communication overhead within the node remains low, which can be attributed to the dedicated communication links, established between each pair of NPUs within the node. However, when $d > 8$, there is a significant increase in communication overhead, which is closely related to the network topology between nodes. Although, in an optimal scenario, each NPU can interconnect with the corresponding rank of NPUs in other nodes via RoCE, the bandwidth is typically several times lower than that of HCCS\footnote{Huawei Cache Coherence System}. This leads to the AR-based TP generally failing to scale effectively across multiple nodes, especially TP is worse than EP when $d = 32$. The A2A-based EP is limited by the inter-node bandwidth; however, it can be integrated with DP to enhance system throughput. Therefore, these complex and difficult-to-optimize parallel strategies present significant challenges for serving MoE models due to issues related to network topology and bandwidth.

The right subfigure of Fig.~\ref{fig:comm_ar_a2a} presents the results for intra-node and inter-node communication with different data sizes. Intra-node communication is based on 4 NPUs within a node, while inter-node communication is based on a single NPU per node across 4 nodes. The results indicate that as the data volume increases, the communication operators within each set initially remain at a relatively low level, after which an inflection point is reached, leading to linear growth. The distinction is that, due to more intra-node bandwidth than inter-node bandwidth, the onset of this inflection point occurs relatively later. Therefore, both in terms of the parallel degree and data size, communication overhead—especially that between nodes—represents a significant bottleneck and a key challenge for the distributed MoE serving system.

\subsection{Opportunities of Decoupling Intra-node Communication \& Inter-node Communication}

We adopt the parallel strategies outlined in the DeepSeek-V3 technical report \cite{DeepSeek-V3}, in which the MoE blocks fully employ EP while retaining the parallel strategy for the Attention blocks. This approach decouples the original EP communication group into intra-node TP groups and inter-node EP groups. We collected the profiling data of the MoE block from a layer of the Decoder in DeepSeek-R1. The results of the Gantt chart in Fig.~\ref{fig:moe_ep_vs_tp_ep_gantt} indicate that while the TP within the nodes introduced AR, it significantly assisted the EP component in sharing a substantial portion of the communication, leading to a significant reduction in the communication overhead of the EP group. Preliminary experimental results indicate that decoupling intra-node and inter-node communication allows for further optimization of communication overhead.

\section{System Design}

We introduce MixServe, a novel automatic distributed serving system that enables efficient deployment of MoE models by TP-EP hybrid parallelism based on a fused AR-A2A communication algorithm. MixServe automatically selects the optimal parallel strategy based on model parameters and network configurations.

\subsection{System Overview}

\begin{figure}[t]
  \centering
  \includegraphics[width=0.45\textwidth]{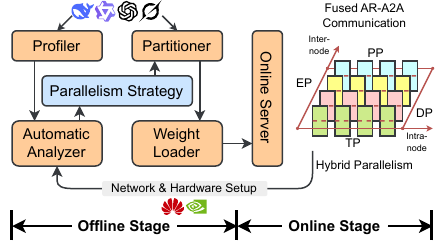}
  \caption{MixServe system overview.}
  \label{fig:ovw}
\end{figure}

Fig.~\ref{fig:ovw} illustrates the system overview of MixServe, which operates in two stages: (i) offline, (ii) online. During the offline stage, MixServe determines the optimal parallelism strategy based on the model's hyperparameters and the configuration of network and hardware resources. During the online stage, MixServe automatically loads and partitions the model weights according to the results of the parallelism strategy analyzed during the offline phase. Additionally, it injects collective communication operators into the model's forward method through the mixed parallel communication groups.

\textbf{Offline Stage}: MixServe first retrieves the model's hyperparameters and presets prompts with varying batch sizes and sequence lengths to obtain profiling data as observations. Subsequently, it uses the configuration of network and hardware resources as input, which includes computational power, as well as intra-node and inter-node network bandwidth and topology, to calculate theoretical values. Both the observations and theoretical values are then input into the analyzer to derive the optimal parallelism strategy. This will provide critical input for the weight loader and partitioner in the online phase.

\textbf{Online Stage}: Based on the optimal parallelism strategy derived from the offline stage, MixServe employs the weight loader to load the corresponding model weight shards through the partitioner. Subsequently, when MixServe initiates the serving service, it initializes the mixed parallel communication group and injects collective communication operators into the appropriate forward method of the MoE models. The serving service manages memory and schedules requests based on the leading vLLM \cite{vLLM} system currently available in the industry.

\subsection{Automatic Analyzer}

\subsubsection{Definition of Parallel Strategies}\label{sec:def}

First of all, in order to facilitate a comprehensive investigation of various parallel strategies, we define a set of context-free grammars to represent the parallel strategies employed by a single Decoder Layer, as follows:

\begin{enumerate}
  \item \textit{strategy} $\longrightarrow$ \textit{Decoder} $\mid$ \textit{Decoder} [PP = \textit{degree}]
  \item \textit{Decoder} $\longrightarrow$ \textit{Attention}, \textit{MoE}
  \item \textit{Attention} $\longrightarrow$ \textit{block}
  \item \textit{MoE} $\longrightarrow$ \textit{block}
  \item \textit{block} $\longrightarrow$ \textit{intra-node} + \textit{inter-node} $\mid$ \textit{parallel}
  \item \textit{intra-node} $\longrightarrow$ \textit{parallel}
  \item \textit{inter-node} $\longrightarrow$ \textit{parallel}
  \item \textit{parallel} $\longrightarrow$ TP $\mid$ EP (DP) = \textit{degree}
  \item \textit{degree} $\longrightarrow 2^k (k \in \mathbb{N})$
\end{enumerate}

The above definition indicates that for each layer of the MoE, the Attention block and the MoE can adopt different parallel strategies. Specifically, the Attention block may utilize TP and DP, while the MoE block may employ TP and EP. It is important to note that we have introduced an additional constraint on the MoE block, namely that DP is typically not considered. This is due to the fact that each expert in the MoE model functions as an independent Multi-Layer Perceptron (MLP), making EP essentially equivalent to DP among the experts. Furthermore, each block may implement different parallel strategies based on the distinct network topologies present both within and between nodes. Furthermore, PP should be applied exclusively between the layers of the Decoder. The previously defined parallelism strategies are confined to a single layer of the Decoder so that they are orthogonal and complementary. The optimal parallel strategy defined herein is the output generated by the automatic analyzer. For example, according to the details presented in the DeepSeek-V3 technical report \cite{DeepSeek-V3}, the parallelism strategy for the prefill phase is TP=4 + DP=8, EP=32.

\subsubsection{Analysis of Collective Communication Operators}\label{sec:comm-ops}

First of all, we conduct a fine-grained analysis of the additional communication overhead resulting from the use of various parallel strategies. Without considering PP, we focus on a single layer of the Decoder in the MoE model.

\begin{table*}[htbp]
  \centering
  \caption{Overhead of collective communication operators.}
  \label{tab:comm-overhead}
  \begin{tabular}{|c|c|cc|c|c|c|c|}
    \hline
    \textbf{Block}                                 &
    \textbf{Strategy}                              &
    \multicolumn{2}{c|}{\makecell{\textbf{Collective} \\ \textbf{Communication}}} &
    \makecell{\textbf{Communication}                  \\ \textbf{per Round}}  &
    \textbf{Algorithm}                             &
    \makecell{\textbf{Rounds of}                      \\ \textbf{Communication}} &
    \makecell{\textbf{Communication}                  \\ \textbf{Domain}}
    \\ \hline
    \multirow{2}{*}{Attention}                     &
    \multirow{2}{*}{TP}                            &
    \multicolumn{1}{c|}{\multirow{2}{*}{AR}}       &
    RS                                             &
    \multirow{2}{*}{$ O(bs \cdot \dfrac{h}{d}) $}  &
    \multirow{2}{*}{Broadcast}                     &
    \multirow{2}{*}{$1$}                           &
    \multirow{2}{*}{Intra-node}
    \\ \cline{4-4}
                                                   &
                                                   &
    \multicolumn{1}{c|}{}                          &
    AG                                             &
                                                   &
                                                   &
                                                   &
    \\ \hline
    \multirow{4}{*}{MoE}                           &
    \multirow{2}{*}{TP}                            &
    \multicolumn{1}{c|}{\multirow{2}{*}{AR}}       &
    RS                                             &
    \multirow{2}{*}{$ O(bs \cdot \dfrac{h}{d}) $}  &
    \multirow{2}{*}{Broadcast}                     &
    \multirow{2}{*}{$1$}                           &
    \multirow{2}{*}{Intra-node}
    \\ \cline{4-4}
                                                   &
                                                   &
    \multicolumn{1}{c|}{}                          &
    AG                                             &
                                                   &
                                                   &
                                                   &
    \\ \cline{2-8}
                                                   &
    \multirow{2}{*}{EP}                            &
    \multicolumn{1}{c|}{\multirow{2}{*}{A2A}}      &
    Dispatch                                       &
    \multirow{2}{*}{$ O(\dfrac{bs}{d} \cdot hk) $} &
    \multirow{2}{*}{Pairwise}                      &
    \multirow{2}{*}{$d-1$}                         &
    \multirow{2}{*}{\makecell{Intra-node or           \\ Inter-node}}
    \\ \cline{4-4}
                                                   &
                                                   &
    \multicolumn{1}{c|}{}                          &
    Combine                                        &
                                                   &
                                                   &
                                                   &
    \\ \hline
  \end{tabular}
\end{table*}

\textbf{AR}: According to the principles of block matrix multiplication, after each rank computes its respective results, it is necessary to sum these results to guarantee the accuracy of the final outcome. This requires performing an AR communication among the ranks. Although the dimension of the tensor $X \in \mathbb{R}^{b \times s \times h}$ \footnote{$b$: batch size, $s$: sequence length, $h$: hidden dimension.}, the AR collective communication operator can be decomposed into RS and AG to reduce the communication volume. According to Table.~\ref{tab:comm-overhead}, the communication volume is $O(bs \cdot \frac{h}{d})$ per round based on Broadcast algorithm. Based on the communication capabilities of modern computational nodes, this process entails full-duplex communication between pairs of devices, which can be accomplished in a single round. Therefore, the overhead of AR communication is theoretically as follows:

\begin{equation}
  \begin{aligned}
    \operatorname{RS}(\mathrm{size}, \mathrm{degree}) & = \operatorname{AG}(\mathrm{size}, \mathrm{degree}) \propto \frac{\mathrm{size}}{\mathrm{degree}}
  \end{aligned}
\end{equation}

\begin{equation}
  \begin{aligned}
    \operatorname{AR}(\mathrm{size}, \mathrm{degree}) & = \operatorname{RS}(\frac{\mathrm{size}}{\mathrm{degree}}, \mathrm{degree}) \\
                                                      & + \operatorname{AG}(\frac{\mathrm{size}}{\mathrm{degree}}, \mathrm{degree})
  \end{aligned}
\end{equation}

\begin{figure}[t]
  \centering
  \subfloat[]{
    \includegraphics[height=0.1875\textheight]{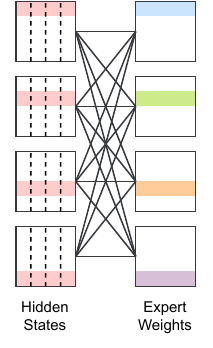}
    \label{fig:DP4-EP4}
  }
  \subfloat[]{
    \includegraphics[height=0.1875\textheight]{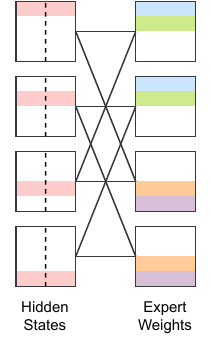}
    \label{fig:DP4-EP2}
  }
  \subfloat[]{
    \includegraphics[height=0.1875\textheight]{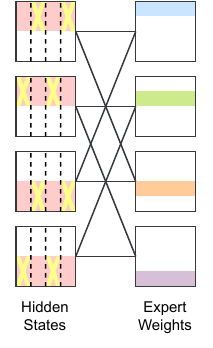}
    \label{fig:DP2-EP4}
  }
  \caption{Trade-off between DP and EP during A2A communication. (a) $d_{\mathrm{DP}} = d_{\mathrm{EP}}$ (Example: DP=4, EP=4). (b) $d_{\mathrm{DP}} > d_{\mathrm{EP}}$ (Example: DP=4, EP=2). (c) $d_{\mathrm{DP}} < d_{\mathrm{EP}}$ (Example: DP=2, EP=4). The yellow cross marks represent the redundancy parts to drop out.}
\end{figure}

\textbf{A2A}: According to the routing mechanism of the MoE models, each token is assigned to its corresponding activated expert for inference. This requires the use of A2A communication to exchange the hidden states among the experts. According to Table~\ref{tab:comm-overhead}, the communication volume is $O(\frac{bs}{d} \cdot hk)$ per round based on Pairwise algorithm, where $k$ denotes the top-$k$ experts activated. The Pairwise algorithm requires $d-1$ rounds to complete the communication process. Therefore, the overhead of A2A communication is theoretically as follows:

\begin{equation}
  \operatorname{A2A}(\mathrm{size}, \mathrm{degree}) \propto \frac{\mathrm{size}}{\mathrm{degree}} \times (\mathrm{degree}-1)
\end{equation}

% \textbf{PP \& P2P}: The PP communication, which is orthogonal to TP, typically facilitates point-to-point (P2P) communication across nodes between corresponding ranks. Furthermore, within the same TP group, data can be sent in shards based on $d_{\mathrm{TP}}$ due to the identical hidden states, followed by an AG communication at the target node.

% \begin{equation}
%   \operatorname{P2P}(\mathrm{size}) \propto \mathrm{size}
% \end{equation}

\subsubsection{Trade-off between DP and EP}\label{sec:trade-off}

Based on the formal definition of parallel strategies presented in \S\ref{sec:def}, an important issue is the integration of DP from the Attention block and EP from the MoE block. Based on the relationships between the parallel degrees, three distinct cases can be identified as follows:

\begin{itemize}
  \item $d_{\mathrm{DP}} = d_{\mathrm{EP}}$: This case represents the most balanced and easily implementable situation, in which the DP rank of the Attention block corresponds one-to-one with the EP rank of the MoE block. In this case, all devices within the communication group engage in A2A communication, as shown in Fig.~\ref{fig:DP4-EP4}.
  \item $d_{\mathrm{DP}} > d_{\mathrm{EP}}$: A smaller $d_{\mathrm{EP}}$ results in redundancy in expert weights, incurring additional memory overhead to enhance DP degree and consequently improve system throughput. In this case, a total of $\frac{d_{\mathrm{DP}}}{d_{\mathrm{EP}}}$ communication groups perform A2A communication in parallel, with each group comprising $d_{\mathrm{EP}}$ devices, as shown in Fig.~\ref{fig:DP4-EP2}.
  \item $d_{\mathrm{DP}} < d_{\mathrm{EP}}$: A smaller $d_{\mathrm{DP}}$ results in redundancy in hidden states and lower throughput; however, this is mitigated by an effective dropping strategy that reduces communication overhead. In this case, a total of $\frac{d_{\mathrm{EP}}}{d_{\mathrm{DP}}}$ communication groups perform A2A communication in parallel, with each group comprising $d_{\mathrm{DP}}$ devices, as shown in Fig.~\ref{fig:DP2-EP4}.
\end{itemize}

Based on the aforementioned analysis, MixServe will automatically manage trade-offs between DP and EP, considering the specified latency and throughput requirements while adhering to memory constraints.

\subsubsection{Token Generation Latency}\label{sec:token_latency}

Aside from embedding and sampling, the latency associated with generating a single token consists of three components: computational latency, communication latency, and an additional queuing latency induced by request contention at the serving system.

\textbf{Computational Latency}: MixServe analyzes and predicts computational cost under hybrid parallelism. The computational latency of each rank can be expressed as follows:

\begin{equation}
  \tau(d_{\mathrm{TP}}, d_{\mathrm{EP}}, d_{\mathrm{DP}}) \propto \frac{\Psi}{d_{\mathrm{TP}} \cdot d_{\mathrm{EP}}} \cdot \frac{b}{d_{\mathrm{DP}}} \cdot sh
\end{equation}

\begin{figure*}[t]
  \centering
  \subfloat[]{
    \includegraphics[height=0.175\textheight]{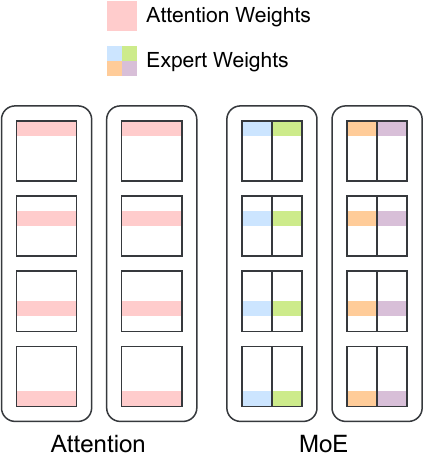}
    \label{fig:MixPrt}
  }
  \subfloat[]{
    \includegraphics[height=0.175\textheight]{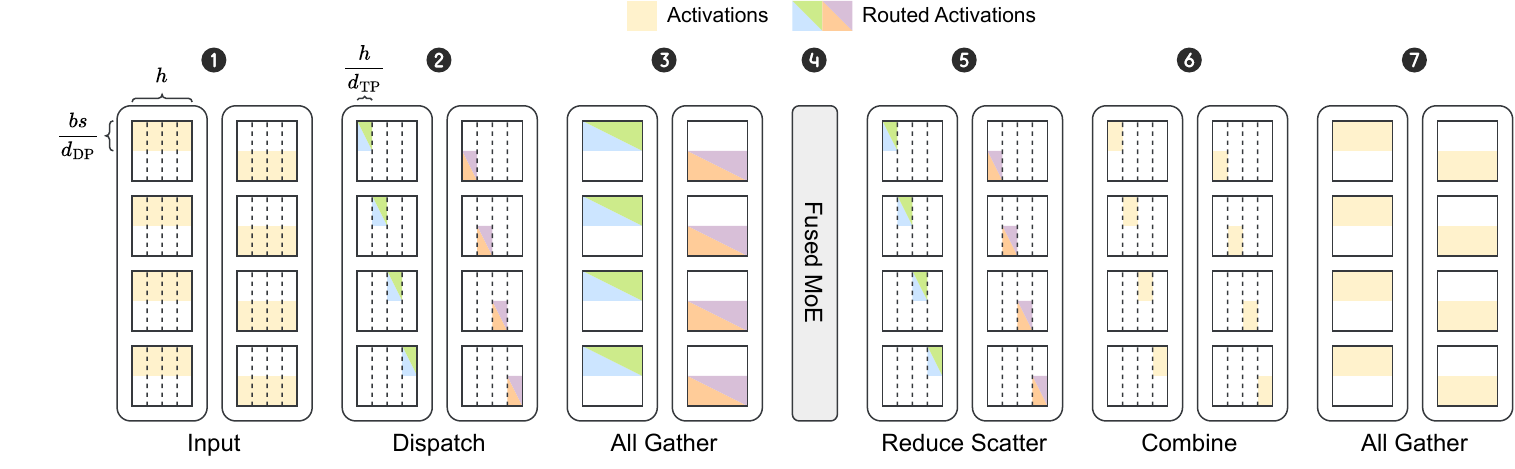}
    \label{fig:MixInfr}
  }
  \caption{An example of TP-EP hybrid parallelism. Assume that we have a 2-node cluster (\textit{i.e.} $n_{\mathrm{node}} = d_{\mathrm{DP}} = 2$) with 4 NPUs each (\textit{i.e.} $n_{\mathrm{proc}} = d_{\mathrm{TP}} = 4$). (a) MoE weights partition. For Attention blocks, model weights are partitioned by intra-node TP and by inter-node DP. For MoE blocks, model weights are partitioned by intra-node TP and by inter-node EP. (b) Distributed MoE inference. Activations are partitioned by inter-node DP with dimension of batch. At the output of MoE blocks, activations are communicated by RS, combine A2A and AG operators.}
  \label{fig:Mix}
\end{figure*}

\textbf{Communication Latency}: According to the analysis in \S\ref{sec:comm-ops} and \S\ref{sec:trade-off}, the communication latency of each rank can be expressed as follows:

\begin{equation}
  \begin{aligned}
    \   & \lambda(d_{\mathrm{TP}}, d_{\mathrm{EP}}, d_{\mathrm{DP}})                                                                                                                                                                     \\
    =\  & 2 \times \operatorname{AR}(\frac{b}{d_{\mathrm{DP}}} \cdot sh, d_{\mathrm{TP}})                                                                                                                                                \\
    +\  & 2 \times \begin{cases}
                     \operatorname{A2A}(\frac{b}{d_{\mathrm{DP}}} \cdot shk, d_{\mathrm{EP}}) & \text{if } d_{\mathrm{DP}} \geq d_{\mathrm{EP}} \\
                     \operatorname{A2A}(\frac{b}{d_{\mathrm{EP}}} \cdot shk, d_{\mathrm{DP}}) & \text{else }
                   \end{cases}
  \end{aligned}
  \label{eq:comm-latency}
\end{equation}

Notice that when $d_{\mathrm{DP}} < d_{\mathrm{EP}}$, the hidden states exhibits $\frac{d_{\mathrm{EP}}}{d_{\mathrm{DP}}}$ times redundancy within DP groups, as shown in Fig.~\ref{fig:DP2-EP4} illustrated in \S\ref{sec:trade-off}. In Eq.~\eqref{eq:comm-latency}, the batch size is specified as $\frac{b}{d_{\mathrm{EP}}}$ because $\frac{b}{d_{\mathrm{EP}}} = \frac{b}{d_{\mathrm{DP}}} / \frac{d_{\mathrm{EP}}}{d_{\mathrm{DP}}}$.

\textbf{Service Latency per Token}: Combining computation and communication, the service latency for generating one token through an $l$-layer Decoder is

\begin{equation}
  \begin{aligned}
    \Delta t_{\mathrm{svc}} & = l[\tau(d_{\mathrm{TP}}, d_{\mathrm{EP}}, d_{\mathrm{DP}}) + \lambda(d_{\mathrm{TP}}, d_{\mathrm{EP}}, d_{\mathrm{DP}})] \\
                            & + (d_{\mathrm{PP}} - 1) \cdot \operatorname{P2P}(\frac{b}{d_{\mathrm{DP}}} \cdot sh)
  \end{aligned}
\end{equation}

Here, $\operatorname{P2P}$ denotes the point-to-point (P2P) communication latency induced by PP.

\textbf{Queuing Latency}: In practical online serving, token generation requests arrive stochastically and may queue before being served. We model the serving system as a queue with arrival rate $\lambda_a$ (tokens per second) and service rate $\mu = 1 / \Delta t_{\mathrm{svc}}$.

For analytical tractability, we adopt an M/M/1 approximation. Under the stability condition $\rho = \lambda_a / \mu < 1$, the expected queuing delay is given by

\begin{equation}
  W_q = \frac{\rho}{\mu (1 - \rho)} = \frac{\lambda_a}{\mu (\mu - \lambda_a)}
\end{equation}

This term captures the contention-induced waiting time caused by concurrent requests, which becomes significant as system utilization approaches saturation.

\textbf{Constraints}: The constraints primarily stem from the limitations imposed by NPU memory, which encompasses two components: model weights and the K/V cache. Assuming that the maximum NPU memory for a single NPU is $M$, it is approximated that the model weights $\Psi$ are primarily composed of Attention blocks and MoE blocks, with a total of $l$ layers in the Decoder.

\begin{equation}
  \frac{\Psi_{\mathrm{Attn}}}{d_{\mathrm{TP}}} +
  \frac{\Psi_{\mathrm{MoE}}}{d_{\mathrm{EP}}d_{\mathrm{TP}}} +
  2bsh \cdot \frac{l}{d_{\mathrm{PP}}} < M
\end{equation}

\subsubsection{Performance Indicators}

In existing evaluations, performance indicators such as TTFT, ITL, and throughput are typically defined based on empirical measurements collected from the serving system. While these observed metrics faithfully reflect end-to-end behavior, they conflate architectural factors with workload-dependent effects. To complement empirical evaluation, we further introduce theoretically estimated performance indicators derived from the latency model in \S\ref{sec:token_latency}, enabling principled analysis and optimization.

\textbf{Time to First Token (TTFT)}: TTFT is defined as the time taken to generate the first token after receiving a request, which reflects the performance of the prefill stage. Under our queuing-aware latency model, TTFT consists of (i) the queuing delay experienced before service and (ii) the service latency required to generate the first token. Importantly, first-token generation differs from steady-state decoding in that the full prompt of length $L_{\mathrm{in}}$ must be processed and the KV cache initialized. Therefore, the theoretically estimated TTFT is defined as:

\begin{equation}
  \operatorname{TTFT} = W_q + \Delta t_{\mathrm{svc}}^{\mathrm{prf}} = W_q + \Delta t_{\mathrm{svc}} \big|_{s=L_{\mathrm{in}}}
\end{equation}

\textbf{Inter-Token Latency (ITL)}: ITL is defined as the average time interval between the generation of two consecutive tokens, which reflects the performance of the decode stage. In this phase, previously computed keys and values are reused via the KV cache, and each iteration only processes a single newly generated token. As a result, the theoretical ITL corresponds to the steady-state per-token service latency:

\begin{equation}
  \operatorname{ITL} = \Delta t_{\mathrm{svc}}^{\mathrm{dec}} =  \Delta t_{\mathrm{svc}} \big|_{s=1}
\end{equation}

\textbf{Throughput}: We model throughput at the service level by jointly accounting for both the prefill stage and the steady-state decoding stage. For a request with input length $L_{\mathrm{in}}$ and output length $L_{\mathrm{out}}$, the expected service time is:

\begin{equation}
  \Theta = \frac{L_{\mathrm{in}} + L_{\mathrm{out}}}{W_q + \Delta t_{\mathrm{svc}}^{\mathrm{prf}} + L_{\mathrm{out}} \cdot \Delta t_{\mathrm{svc}}^{\mathrm{dec}}}
\end{equation}

\subsection{Hybrid TP-EP Partitioner}\label{sec:mix-part}

\begin{figure}[t]
  \centering
  \includegraphics[width=0.475\textwidth]{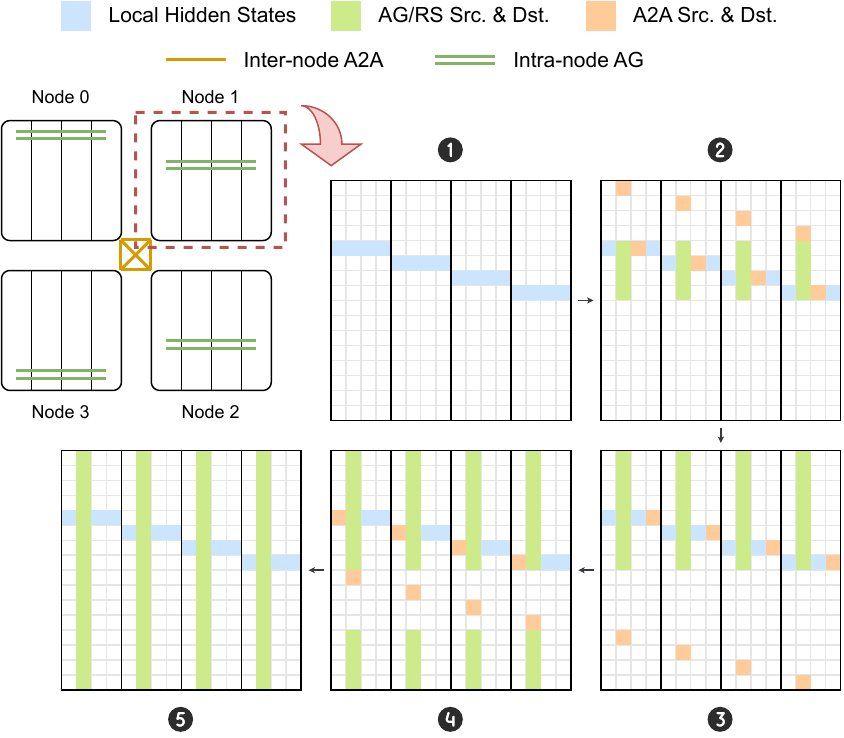}
  \caption{An example of fused AR-A2A communication algorithm. Assume that we have a 4-node cluster with 4 GPUs each. Steps 1-5 illustrates how hidden states are synchronized by intra-node TP and inter-node EP.}
  \label{fig:Fused-AR-A2A}
\end{figure}

\subsubsection{Hybrid TP-EP Design}

Fig.~\ref{fig:Mix} illustrates a simple design of hybrid TP-EP. In this example, we assume that there is a 2-node cluster (\textit{i.e.} $n_{\mathrm{node}} = d_{\mathrm{DP}} = 2$) with 4 NPUs each (\textit{i.e.} $n_{\mathrm{proc}} = d_{\mathrm{TP}} = 4$). The Attention blocks are partitioned by intra-node TP and by inter-node DP, while the MoE blocks are partitioned by intra-node TP and by inter-node EP. Considering that the hybrid TP-EP involves communication between two distinct communication groups, MixServe has implemented fine-grained optimizations on the communication process. Specifically, MixServe decouples the AR of the TP group into RS and AG, and reorganizes the A2A of the EP group, ultimately forming the RS-A2A-AG communication process. Correspondingly, MixServe injects the specified parallel strategy into the model's weight loader through the partitioner.

\subsubsection{Theoretical Analysis of Communication} Existing serving systems commonly utilize EP entirely within the MoE blocks, while TP is used in the Attention blocks in intra-node. Assuming a cluster of $n_{\mathrm{node}}$ nodes, each with $n_{\mathrm{proc}}$ NPUs, the parallel strategy is defined as $\mathrm{TP} = n_{\mathrm{proc}} + \mathrm{DP} = n_{\mathrm{node}}$, $\mathrm{EP} = n_{\mathrm{node}} \cdot n_{\mathrm{proc}}$. Thus, communication overhead of a Docoder layer is as follows:

\begin{equation}
  \begin{aligned}
    \lambda_{\mathrm{EP}} & = \operatorname{AR} (bsh, n_{\mathrm{proc}})            \\
                          & + 2 \times \operatorname{A2A} (bshk, n_{\mathrm{node}}) \\
  \end{aligned}
\end{equation}

The hybrid TP-EP parallelism decouples and reorganizes AR and A2A, resulting in a reduction of the per-unit communication volume of the Combine stage, as well as a decrease in communication scale both intra-nodes and inter-nodes. Considering the impact of network topology and the interconnection methods between nodes and NPUs, pure TP or EP is often constrained by inter-node bandwidth, which prevents the optimization of the overall communication group's efficiency. When $n_{\mathrm{proc}} = d_{\mathrm{TP}}$ and $n_{\mathrm{node}} = d_{\mathrm{EP}}$, the TP group and EP are precisely allocated to intra-nodes and inter-nodes. The parallel strategy of MixServe is defined as $\mathrm{TP} = n_{\mathrm{proc}} + \mathrm{DP} = n_{\mathrm{node}}$, $\mathrm{TP} = n_{\mathrm{proc}} + \mathrm{EP} = n_{\mathrm{node}}$. The communication overhead at this stage is as follows:

\begin{equation}
  \begin{aligned}
    \lambda_{\mathrm{mix}} & =  \operatorname{AR} (bsh, n_{\mathrm{proc}}) + \operatorname{AG} (\frac{bshk}{n_{\mathrm{proc}}}, n_{\mathrm{proc}}) \\
                           & + 2 \times \operatorname{A2A} (\frac{bshk}{n_{\mathrm{proc}}}, n_{\mathrm{node}})
  \end{aligned}
\end{equation}

Certainly, MixServe is not restricted to merely identifying $n_{\mathrm{proc}} = d_{\mathrm{TP}}$ and $n_{\mathrm{node}} = d_{\mathrm{EP}}$ as the optimal parallel strategy; instead, it conducts theoretical analyses and predictions for all parallel strategies that satisfy $n_{\mathrm{proc}} \cdot n_{\mathrm{node}} = d_{\mathrm{TP}} \cdot d_{\mathrm{EP}}$. Following the completion of the parallel strategy decision for MoE blocks, the parallel strategy for Attention blocks will be optimized in conjunction with the theoretical analysis presented in \S\ref{sec:trade-off}. Consequently, the partitioner will output the optimal parallel strategy.

\subsection{Fused AR-A2A Communication Algorithm}\label{sec:fused-ar-a2a}

Building upon the hybrid TP-EP parallelism, we design the fused AR-A2A communication algorithm by employing the principle of mutual overlapping of intra-node and inter-node communication, guided by the computational dependency relationships.

\begin{figure}[t]
  \centering
  \subfloat[Fused RS-Combine Gantt Chart]{
    \includegraphics[width=0.45\textwidth]{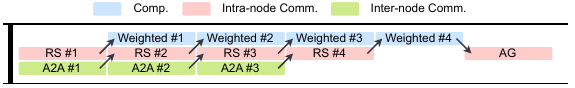}
    \label{fig:Fused-RS-Combine-Gantt}
  } \\
  \subfloat[Fused AG-Dispatch Gantt Chart]{
    \includegraphics[width=0.45\textwidth]{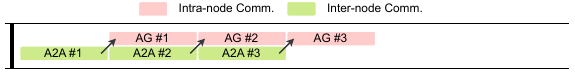}
    \label{fig:Fused-AG-Dispatch-Gantt}
  }
  \caption{Gantt chart of fused AR-A2A communication algorithm. (a) Fused RS-Combine communication algorithm. (b) Fused AG-Dispatch communication algorithm. Both of them facilitate the overlapping of intra-node and inter-node communication.}
  \label{fig:Fused-AR-A2A-Gantt}
\end{figure}

Fig.~\ref{fig:Fused-AR-A2A} illustrates the overall process of the fused AR-A2A communication algorithm. Steps 1-5 demonstrate how hidden states are synchronized by intra-node TP and inter-node EP. Initially, each rank only possesses a partition of the hidden states (the blue segment). Subsequently, intra-node AG/RS communication (the light green segment) and inter-node A2A communication (the orange segment) are performed to exchange the hidden states asynchronously step by step. Finally, each rank acquires the complete hidden states (the green segment in step 5) after the communication process concludes.

\subsubsection{Fused RS-Combine Communication Algorithm}

\begin{algorithm}[t]
  \caption{Fused RS-Combine Pairwise Communication}\label{alg:Fused-RS-Combine}
  \begin{algorithmic}[1]
    \Require An $n$-node cluster with $m$ GPUs/NPUs per node; an input tensor $X \in \mathbb{R}^{\frac{bs}{d_{\mathrm{EP}}} \times h}$ per node; global rank $r$
    \Ensure An output tensor $Y \in \mathbb{R}^{\frac{b}{d_{\mathrm{DP}}} \times s \times h}$ each node

    \State $Y \gets \mathtt{empty} (\frac{b}{d_{\mathrm{DP}}}, s, h)$
    \State $[X_1, X_2, \cdots, X_m] \gets \mathtt{split} (X, m, -1)$ \Comment{Split $X$ into $m$ parts along the hidden dimension (the same below)}
    \State $[Y_1, Y_2, \cdots, Y_m] \gets \mathtt{split} (Y, m, -1)$
    \State $r_{\mathrm{TP}} \gets r \bmod m$ \Comment{Compute TP rank}

    \State Initialize tensor list $[S_1, S_2, \cdots, S_n]$
    \State $S_1 \gets X_{r_{\mathrm{TP}}}$ \Comment{Stage local tensor $X_{r_{\mathrm{TP}}}$}

    \For{$i \gets 1$ to $n-1$} \textbf{async}
    \State $r_{\mathrm{to}} \gets (r_{\mathrm{TP}} + im) \bmod mn$
    \State $\mathtt{isend}(X_{r_{\mathrm{TP}}}, r_{\mathrm{to}})$ \Comment{Send $X_{r_{\mathrm{TP}}}$ to the same TP rank of the next $i$-step node asynchronously}
    \State $r_{\mathrm{from}} \gets (r_{\mathrm{TP}} - im) \bmod mn$
    \State $S_{i+1} \gets \mathtt{irecv}(r_{\mathrm{from}})$ \Comment{Receive $X_{r_{\mathrm{TP}}}$ from the same TP rank of the previous $i$-step node asynchronously}
    \EndFor \Comment{Inter-node A2A pairwise communication}

    \For{$i \gets 1$ to $n$} \textbf{async}
    \State $S_i \gets$ \textbf{await} $\mathtt{reduce\_scatter} (S_i, \mathrm{TP\ group})$
    \State $Y_i \gets Y_i + \mathtt{topk\_weights} (S_i)$
    \EndFor \Comment{Intra-node AR communication}

    \State $Y \gets \mathtt{all\_gather} (Y_{r_{\mathrm{TP}}}, \mathrm{TP\ group})$
  \end{algorithmic}
\end{algorithm}

Fig.~\ref{fig:Fused-RS-Combine-Gantt} and Alg.~\ref{alg:Fused-RS-Combine} illustrates the fused RS-Combine communication algorithm. The algorithm is designed to optimize the communication process by overlapping intra-node and inter-node communication. The key steps are as follows: (1) Intra-node RS, (2) Inter-node A2A, (3) Intra-node AG.

Initially, the hidden states at each rank within the node engage in one round of RS communication, temporarily storing the results after weighting them with the top-k weights. Concurrently, a round of communication between nodes is executed using the Pairwise algorithm, enabling each node to acquire the corresponding hidden states as input for the subsequent iteration. Upon completion of the Pairwise algorithm, all weighted results are ultimately combined through AG within the nodes. The Gantt chart in Fig.~\ref{fig:Fused-AR-A2A-Gantt} illustrates the overlapping of communication processes, where the RS and A2A communication are executed concurrently, followed by the AG operation.

In summary, the algorithm necessitates $n_{\mathrm{node}} - 1$ rounds of communication between nodes and $n_{\mathrm{node}}$ rounds of communication within each node. The asynchronous mechanism facilitates overlapping communication both within and across nodes, resulting in a time complexity of $O(n_{\mathrm{node}})$. Furthermore, the algorithm necessitates the allocation of additional temporary storage space for each rank, corresponding in size to the output. Consequently, the space complexity is $O(bsh \cdot n_{\mathrm{proc}})$ in total. The fused RS-Combine Pairwise communication algorithm we present ingeniously incorporates an overlapping mechanism for communication both within and between nodes, effectively trading off space for time. This approach significantly reduces the communication overhead associated with inference in the MoE models.

\subsubsection{Fused AG-Dispatch Communication Algorithm}

Similarly, it is precisely because the hidden states are replicated in the MoE TP group that they can be sharded within the MoE TP group. The sharding further minimizes the inter-node Dispatch communication overhead, requiring only the addition of extra intra-node AG communication, analogous to Megatron-based PP. On this basis, it is possible to allow for the overlapping of the intra-node AG communication and inter-node Dispatch communication, in a manner analogous to the Fused RS-Combine algorithm.

\begin{algorithm}[t]
  \caption{Fused AG-Dispatch Pairwise Communication}\label{alg:Fused-AG-Dispatch}
  \begin{algorithmic}[1]
    \Require An $n$-node cluster with $m$ GPUs/NPUs per node; an input tensor $X \in \mathbb{R}^{\frac{b}{d_{\mathrm{DP}}} \times s \times h}$ per node; global rank $r$
    \Ensure An output tensor $Y \in \mathbb{R}^{\frac{bs}{d_{\mathrm{EP}}} \times h}$ per node

    \State $Y \gets \mathtt{empty} (\frac{bs}{d_{\mathrm{EP}}}, h)$
    \State $[X_1, X_2, \cdots, X_m] \gets \mathtt{split} (X, m, -1)$
    \State $[Y_{11}, Y_{12}, \cdots, Y_{mn}] \gets \mathtt{split} (Y, [m, n], [0, 1])$ \Comment{Split $Y$ into $m \times n$ parts along the token and hidden dimension}
    \State $r_{\mathrm{TP}} \gets r \bmod m$
    \State $[X_{r_{\mathrm{TP}} 1}, X_{r_{\mathrm{TP}} 2}, \cdots, X_{r_{\mathrm{TP}} n}] \gets \mathtt{route} (X_{r_{\mathrm{TP}}})$ \Comment{Calculate the expert map on local TP rank only}

    \For{$i \gets 1$ to $n-1$} \textbf{async}
    \State $r_{\mathrm{to}} \gets (r_{\mathrm{TP}} + im) \bmod mn$
    \State $\mathtt{isend}(X_{r_{\mathrm{TP}} i}, r_{\mathrm{to}})$
    \State $r_{\mathrm{from}} \gets (r_{\mathrm{TP}} - im) \bmod mn$
    \State $Y_{r_{\mathrm{TP}} r_{\mathrm{from}}} \gets \mathtt{irecv}(r_{\mathrm{from}})$
    \EndFor

    \For{$i \gets 1$ to $n$} \textbf{async}
    \State $Y_{:i} \gets$ \textbf{await} $\mathtt{all\_gather} (Y_{r_{\mathrm{TP}} i}, \mathrm{TP\ group})$
    \EndFor
  \end{algorithmic}
\end{algorithm}

Fig.~\ref{fig:Fused-AG-Dispatch-Gantt} shows the Gantt chart of the fused AG-Dispatch communication algorithm. Apart from the pairwise communication in the first round and the AG communication in the last round, the intra-node and inter-node communication during the remaining rounds can overlap with one another. Alg.~\ref{alg:Fused-AG-Dispatch} describes the detailed communication schedule. In contrast to Alg.~\ref{alg:Fused-RS-Combine}, the total number of communication rounds both within and between nodes is $n_{\mathrm{node}} - 1$, as the local shards in the TP group and EP group do not require communication. Therefore, the time complexity of the algorithm is $O(n_{\mathrm{node}})$ and the space complexity is $O(1)$, respectively.

\section{Evaluation}

\subsection{Experimental Setup}

% \textbf{Hardware and Network}: We conduct our experiments on a cluster of 4 nodes, each equipped with 8 NVIDIA A800 GPUs (80 GB). The intra-node network is fully-connected via NVLink (400 GB/s), while the inter-node network is connected via InfiniBand (100 Gbps). In detail, each GPU has a connection to a 32-port switch (8 ports usage for each node), thus 4 nodes need 2 switches, and the 2 switches also need to connect each other (remaining 16 ports usage for each switch). As \S\ref{sec:mix-part} describes, we impose a restriction on inter-node communication such that only the GPUs corresponding to the ranks on each node are permitted to communicate, in order to improve network throughput.

\textbf{Hardware and Network}: We conduct our experiments on following clusters:

\begin{itemize}
  \item A cluster of 2 servers with 8 Nvidia H20 GPUs (96 GB) each. The intra-node network is supported by NVLink 4.0 (up to 900 GB/s), while the inter-node network is connected via InfiniBand (400 Gbps).
  \item A cluster of 4 Atlas 800T A2 servers with 8 Ascend 910B NPUs (64 GB) each. The intra-node network is fully-connected via HCCS (up to 480 Gbps), while the inter-node network is connected via RoCE (up to 200 Gbps).
\end{itemize}

\textbf{Implementation}: We implement MixServe based on several serving systems, including vLLM \cite{vLLM} (on the Ascend 910B cluster) and Tutel \cite{Tutel} (on the H20 cluster).

\textbf{Models and Datasets}: To evaluate MixServe, we adopt the following SOTA MoE models: (1) DeepSeek-R1 \cite{DeepSeek-R1}, a 671B-parameter MoE model with 256 routed experts and 1 shared expert, where 37B parameters are activated per token; and (2) Qwen3 \cite{Qwen3}, a 235B-parameter MoE model with 128 experts, with 22B parameters activated per token. We use ShareGPT-V3 \cite{ShareGPT-v3} for benchmark evaluation, which is a large-scale dataset containing 1.2B tokens of human conversations.

\textbf{Baselines}: We compare MixServe with the following baselines: (1) vLLM \cite{vLLM}, which utilizes hybrid TP+PP for LLM serving and hybrid DP+EP for distributed MoE model serving; and (2) Tutel \cite{Tutel}, which employs hybrid TP+EP for distributed MoE model serving. In addition, we also set up different TP degrees (\textit{i.e.} 4 and 8) for comparative experiments. The specific configurations of parallel strategies for baselines are summarized in Table~\ref{tab:config}.

\begin{table}
  \centering
  \caption{Configuration of parallel strategies of baselines.}
  \label{tab:config}
  \begin{tabular}{|c|c|c|}
    \hline
    \multirow{2}{*}{\textbf{Baselines}} & \multicolumn{2}{c|}{\textbf{Parallel Strategies}}               \\ \cline{2-3}
     & \textbf{H20} &  \textbf{Ascend 910B} \\ \hline
    vLLM       & \makecell{TP=8 [PP=2]\\TP=8 + DP=2, EP=16\\TP=4 + DP=4, EP=16}  & \makecell{TP=8 [PP=4]\\TP=8 + DP=4, EP=32\\TP=4 + DP=8, EP=32}   \\ \hline
    Tutel       & \makecell{TP=8 + DP=2, TP=8 + EP=2\\TP=4 + DP=4, TP=4 + EP=4}  & \textit{Not supported}   \\ \hline
  \end{tabular}
\end{table}

\subsection{Performance Evaluation}

\begin{figure}[t]
  \centering
  \subfloat[TTFT]{
    \includegraphics[width=0.45\textwidth]{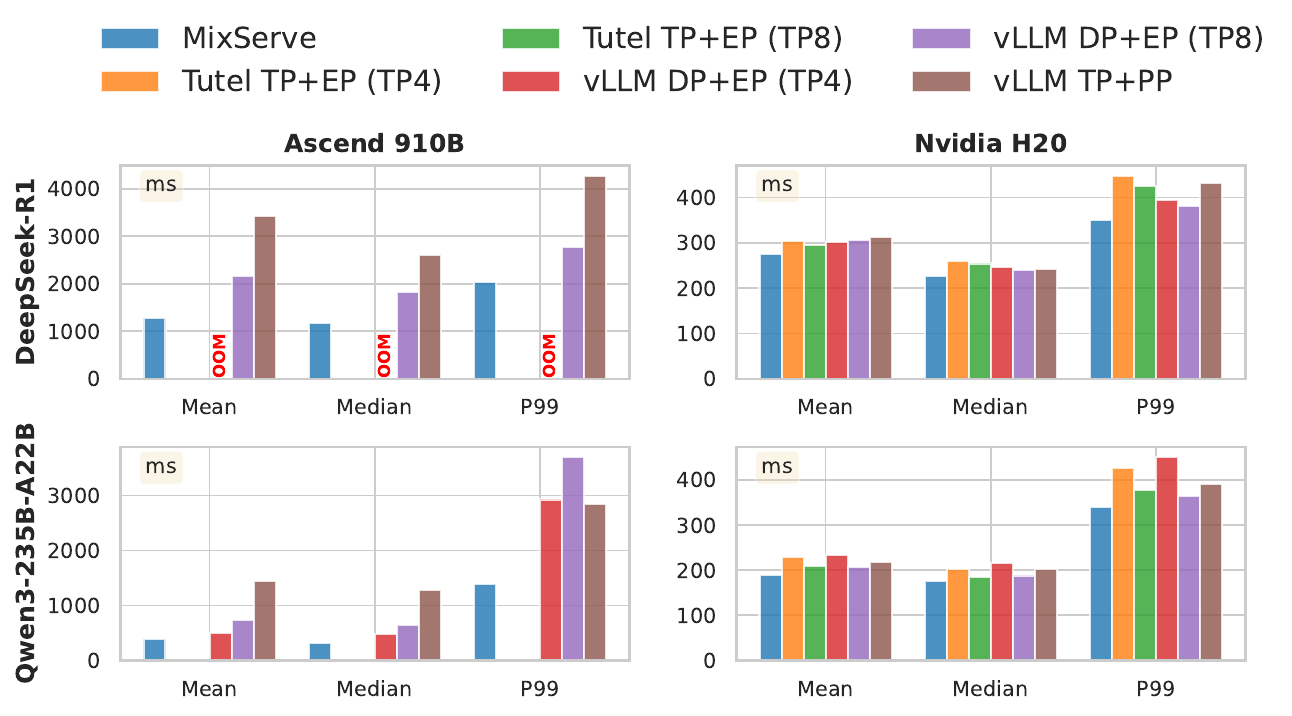}
    \label{fig:TTFT}
  }\\
  \subfloat[ITL]{
    \includegraphics[width=0.45\textwidth]{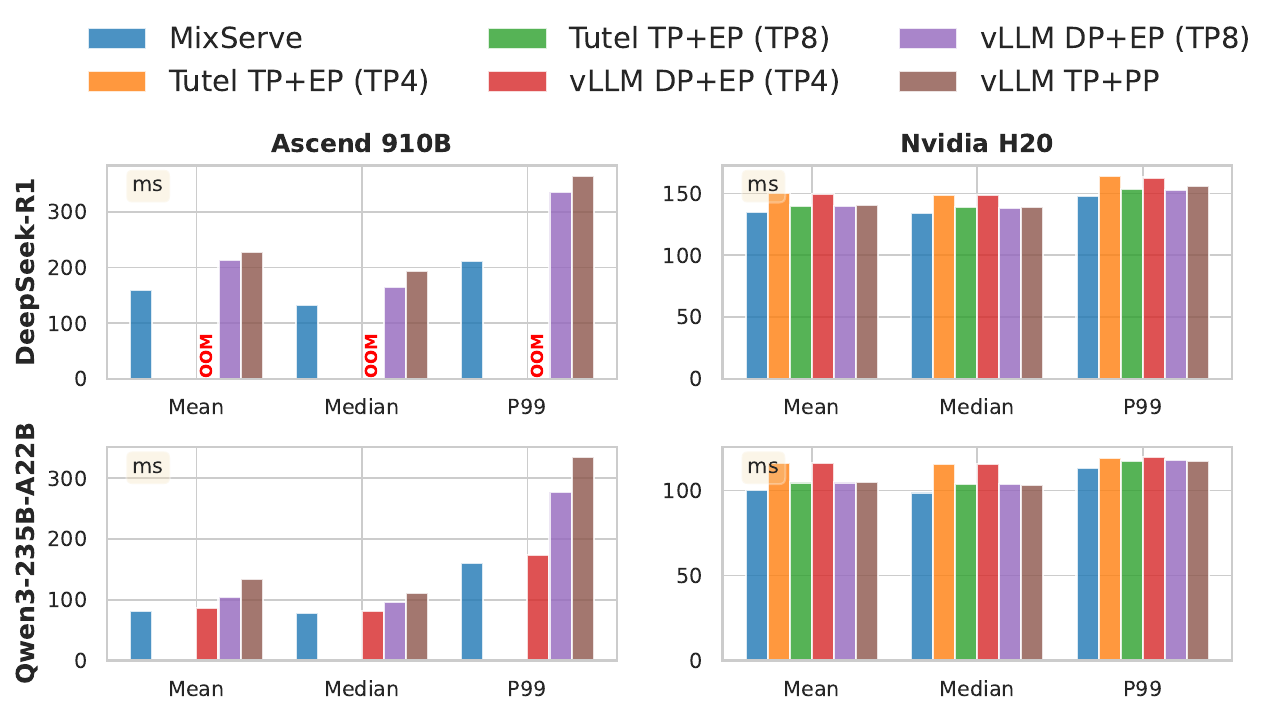}
    \label{fig:ITL}
  }\\
  \subfloat[Throughput]{
    \includegraphics[width=0.45\textwidth]{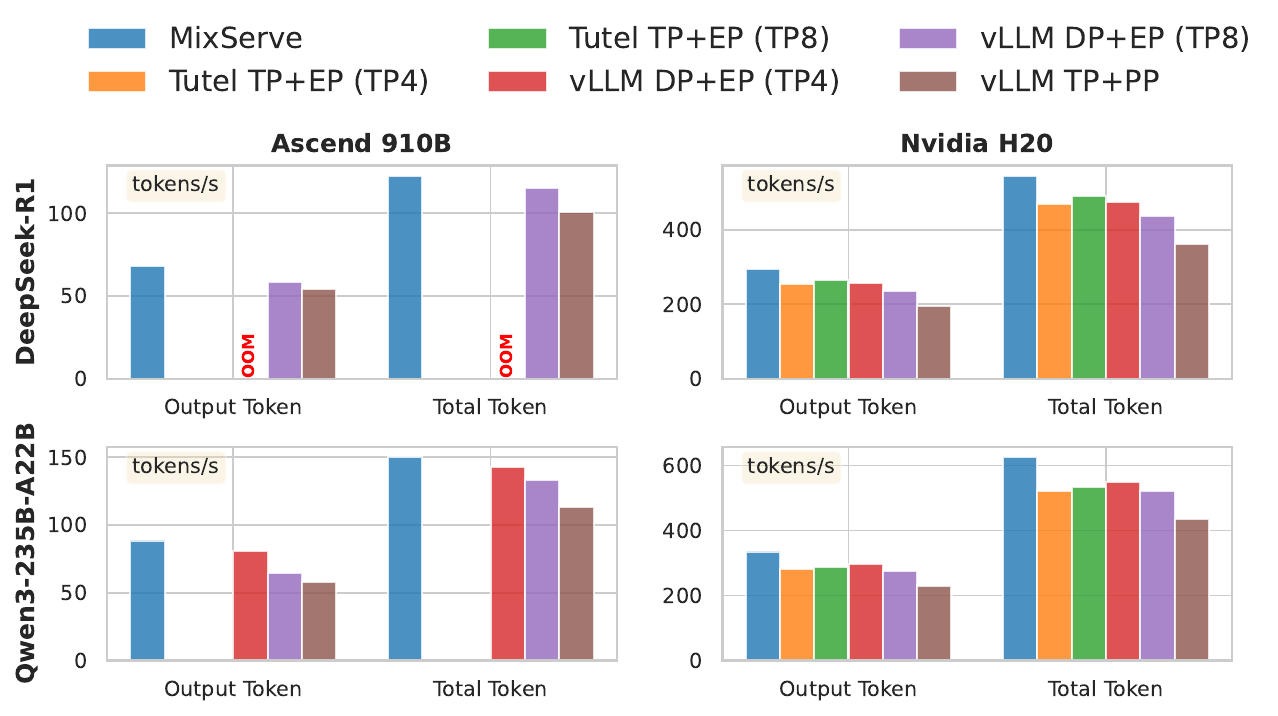}
    \label{fig:Throughput}
  }
  \caption{Performance evaluation of MixServe and baselines. The results are averaged over 10 runs, with error bars representing the standard deviation.}
  \label{fig:Performance}
\end{figure}

We established the range of request rates at 2, 4, and 8 requests per second (req/s), while also defining the maximum batch size as 16 and the maximum sequence length as 4096 tokens.  Fig.~\ref{fig:Performance} shows the performance evaluation of MixServe and baselines across different metrics.

\textbf{TTFT}: Fig.~\ref{fig:TTFT} illustrates that MixServe achieves significantly lower TTFT compared to baselines, indicating faster response times during the prefill stage. Specifically, MixServe achieves $1.08 \times \sim 3.80 \times$ acceleration in TTFT across different configurations and models. On the Ascend 910B cluster, MixServe demonstrates particularly impressive improvements: for DeepSeek-R1, it achieves $2.67 \times$ acceleration compared to vLLM TP+PP and $1.70 \times$ compared to vLLM DP+EP; for Qwen3-235B-A22B, it achieves $3.80 \times$ acceleration compared to vLLM TP+PP and $1.32 \times \sim 1.93 \times$ compared to vLLM DP+EP configurations. On the H20 cluster, MixServe achieves $1.08 \times \sim 1.23 \times$ acceleration compared to various baselines. The experimental results demonstrate that: (1) the hybrid TP-EP parallelism proposed by MixServe effectively reduces TTFT across diverse hardware platforms and model architectures; (2) the overlapping communication between intra-nodes and inter-nodes significantly reduces overall communication overhead, resulting in improved P99\footnote{P99 refers to the 99th percentile latency, which means 99\% of requests are served within this time.} performance for MixServe.

\textbf{ITL}: Fig.~\ref{fig:ITL} shows that MixServe demonstrates lower ITL, indicating faster token generation during the decode stage. The hybrid TP-EP parallelism achieves $1.03 \times \sim 1.66 \times$ acceleration across all evaluated configurations. On the Ascend 910B cluster, MixServe reduces ITL from 227.33ms to 160.06ms ($1.42 \times$ acceleration) for DeepSeek-R1 compared to vLLM TP+PP, and from 134.27ms to 81.1ms ($1.66 \times$ acceleration) for Qwen3-235B-A22B. On the H20 cluster, MixServe achieves $1.03 \times \sim 1.16 \times$ acceleration compared to various baselines. Although the acceleration effect is less pronounced than TTFT due to the smaller communication volume in the decode stage, the consistent improvements demonstrate the effectiveness of the fused AR-A2A communication algorithm.

\textbf{Throughput}: Fig.~\ref{fig:Throughput} illustrates that MixServe achieves substantially higher throughput compared to baselines, allowing it to handle more requests simultaneously and improve overall system efficiency. The total token throughput improvements range from 5.2\% to 50.3\% across different configurations. On the Ascend 910B cluster, MixServe achieves 22.0\% throughput improvement (from 100.61 to 122.72 tokens/s) for DeepSeek-R1 and 32.2\% improvement (from 113.52 to 150.08 tokens/s) for Qwen3-235B-A22B compared to vLLM TP+PP. On the H20 cluster, the improvements are even more substantial: 50.3\% for DeepSeek-R1 (from 362.78 to 545.23 tokens/s) and 43.5\% for Qwen3-235B-A22B (from 435.82 to 625.45 tokens/s) compared to vLLM TP+PP. When compared to other EP-based approaches, MixServe consistently achieves 6.8\% $\sim$ 24.5\% throughput improvements, demonstrating the effectiveness of the automatic parallel strategy selection and fused communication algorithm.

\begin{figure}[t]
  \centering
  \includegraphics[width=0.45\textwidth]{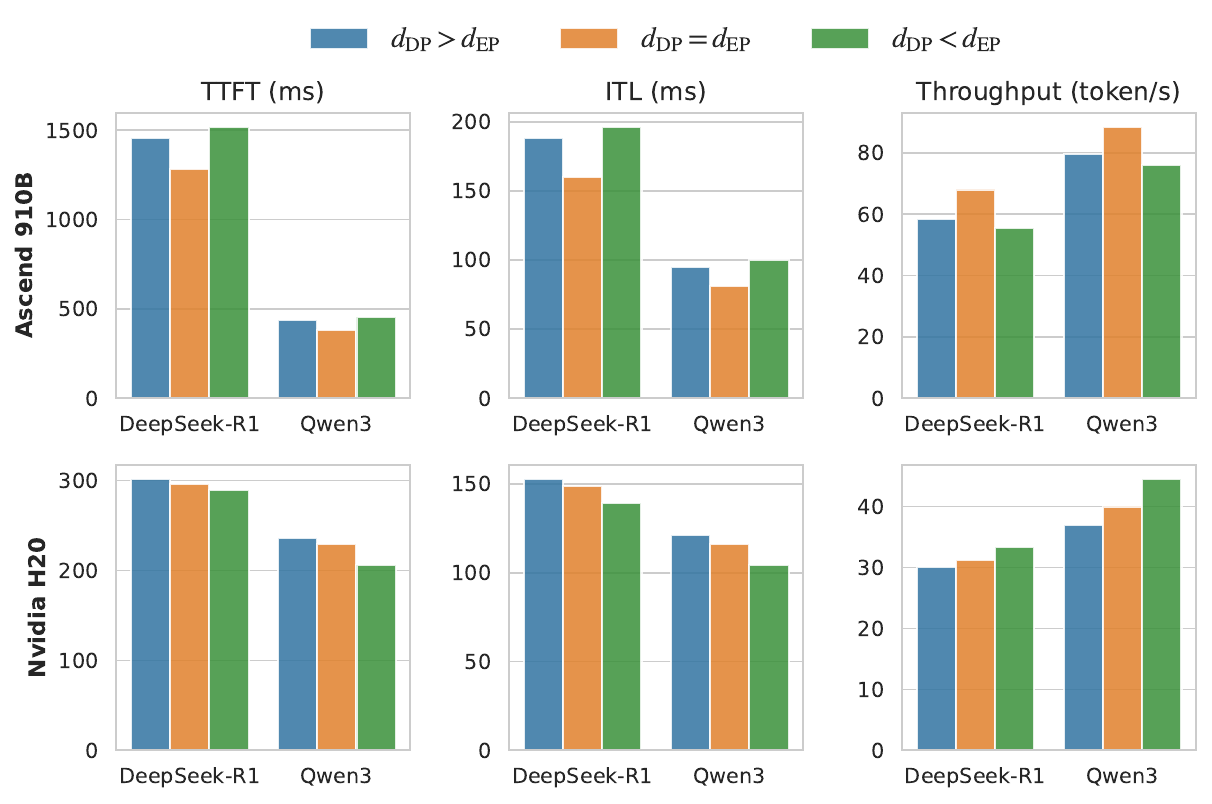}
  \caption{Performance comparison of MixServe with different DP and EP configurations.}
  \label{fig:DP_EP_Comparison}
\end{figure}

\subsection{Ablation Studies}

To better understand the impact of different components in MixServe, we conduct ablation studies by systematically removing or modifying key features.

\subsubsection{Trade-off between DP and EP}

As \S\ref{sec:trade-off} describes, MixServe optimizes $d_{\mathrm{DP}}$ and $d_{\mathrm{EP}}$ by evaluating the modeled communication and computation costs across feasible TP/DP/EP tuples. We study three representative settings: (1) $d_{\mathrm{DP}} = d_{\mathrm{EP}}$ (TP=8 + DP=4, TP=8 + EP=4), (2) $d_{\mathrm{DP}} > d_{\mathrm{EP}}$ (TP=4 + DP=8, TP=8 + EP=4), and (3) $d_{\mathrm{DP}} < d_{\mathrm{EP}}$ (TP=8 + DP=4, TP=4 + EP=8).

Fig.~\ref{fig:DP_EP_Comparison} summarizes the ablation results. On Ascend 910B, the balanced case attains the best latency/throughput for both DeepSeek-R1 and Qwen3 (e.g., 383.14ms TTFT and 150.08 tokens/s throughput for Qwen3 when $d_{\mathrm{DP}} = d_{\mathrm{EP}}$), while skewing towards larger DP or larger EP degrades performance. However, on Nvidia H20, a different ordering holds: $d_{\mathrm{DP}} < d_{\mathrm{EP}}$ yields the lowest TTFT (e.g., 228.99ms for Qwen3) and the highest throughput (40.00 tokens/s). These observations align with our analytical trade-off model—balancing DP and EP minimizes the dominant communication term—so the partitioner automatically selects this configuration under both high-bandwidth NVLink (H20) and RoCE/HCCS (910B) environments. When cluster bandwidth or node count changes, MixServe re-evaluates the cost model and picks the best feasible tuple, ensuring the serving system adapts its parallel strategy to the available network and compute resources.

\subsubsection{Impact of Overlapping Communication}

As \S\ref{sec:fused-ar-a2a} describes, MixServe employs a fused AR-A2A communication algorithm to optimize the communication process. We evaluate the impact of this optimization on performance by whether asynchronous or synchronous communication is used.

\begin{figure}[t]
  \centering
  \subfloat[]{
    \includegraphics[width=0.45\textwidth]{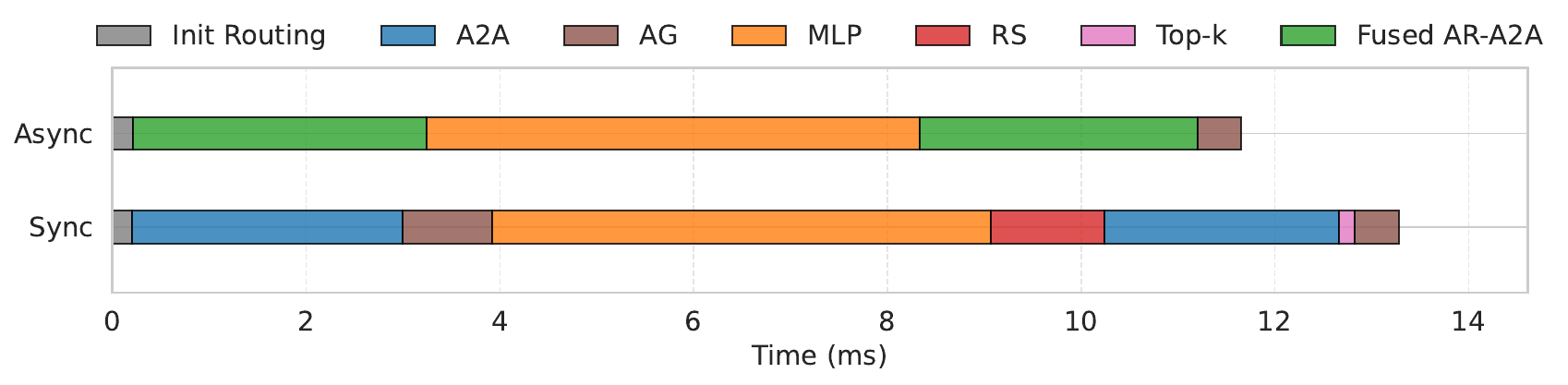}
    \label{fig:Sync_Async_Gantt}
  } \\
  \subfloat[]{
    \includegraphics[width=0.45\textwidth]{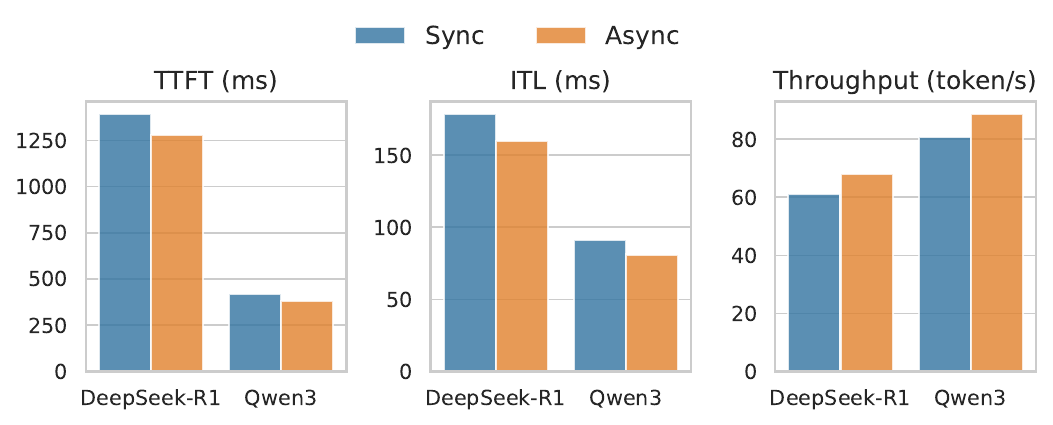}
    \label{fig:Sync_Async_Comparison}
  }
  \caption{Impact of overlapping communication based on fused AR-A2A communication algorithm in MixServe. We evaluate that on the Ascend 910B cluster with DeepSeek-R1. (a) Gantt chart of Sync and Async communication. (b) Performance comparison of Sync and Async communication.}
  \label{fig:Sync_Async_Impact}
\end{figure}

Fig.~\ref{fig:Sync_Async_Gantt} shows the Gantt chart of synchronous and asynchronous communication, where the asynchronous communication allows for overlapping of intra-node and inter-node communication. Specifically, the fused AG-Dispatch communication algorithm overlaps inter-node Dispatch communication with intra-node AG communication, while the fused RS-Combine algorithm overlaps inter-node Combine communication, intra-node RS communication, and the computation of top-$k$ weights. The Gantt chart indicates that the asynchronous fused AR-A2A demonstrates a performance improvement compared to the total latencies of the synchronous operators, which is approximately slightly greater than inter-node communication overhead.

Fig.~\ref{fig:Sync_Async_Comparison} shows the performance comparison of synchronous and asynchronous communication. The results indicate that the asynchronous communication significantly reduces the overall latency, leading to improved TTFT and ITL. The throughput also increases due to the reduced communication overhead. This ablation study demonstrates the effectiveness of overlapping communication in enhancing performance.

\section{Related Work}

\subsection{Distributed MoE}

In the early stages of research, various distributed methods facilitated parallel training of MoE models to improve throughput and efficiency. GShard \cite{GShard} pioneering the use of all-to-all communication for large-scale sparsity. Subsequent frameworks like DeepSpeed-MoE \cite{DeepSpeed-MoE} and Tutel \cite{Tutel} refined this via hybrid data/expert parallelism and fused kernels to enhance memory efficiency and multi-node scalability. To further mitigate communication bottlenecks, SmartMoE \cite{SmartMoE} introduced dynamic strategy selection, while Lina \cite{Lina} optimized interleaved all-to-all operators. More recently, the field has moved toward high-dimensional parallelism; notably, MoE Parallel Folding \cite{liu2025moeparallelfoldingheterogeneous} utilizes Megatron-Core \cite{Megatron-Core} to integrate TP, EP, DP, PP, and context parallelism (CP) into a unified 5D scheme for heterogeneous clusters.

Our work leverages numerous methods and concepts from the training of distributed MoE models, focusing on their application in distributed MoE model serving.

\subsection{Distributed LLM Serving}

Distributed serving systems prioritize maximizing throughput and minimizing latency for online inference. Early optimizations focused on request scheduling: Orca \cite{Orca} pioneered iteration-level scheduling and selective batching, while Llumnix \cite{Llumnix} introduced dynamic resource allocation based on workload characteristics. Regarding parallelism, Alpa \cite{Alpa} and AlpaServe \cite{AlpaServe} explored the synergy between intra/inter-operator parallelism and model multiplexing. To further optimize the distinct phases of inference, DistServe \cite{DistServe} proposed prefill/decode (P/D) disaggregation, and Sarathi-Serve \cite{Sarathi-Serve} utilized chunked-prefills with stall-free scheduling to balance throughput-latency trade-offs. Most recently, MegaScale-Infer \cite{MegaScale-Infer} extended disaggregation to Attention and MoE blocks, leveraging ping-pong pipeline parallelism to hide communication overhead and maximize GPU utilization.

Our work focuses on parallel strategies and communication optimization, and can be effectively incorporated with various optimization methods of existing LLM serving systems, such as request scheduling, P/D disaggregation, etc.

\section{Conclusion}

We introduce MixServe, a novel automatic distributed serving system that for efficient deployment of MoE models by hybrid TP-EP based on fused AR-A2A communication algorithm. MixServe automatically selects the optimal parallel strategy based on model parameters and network configurations. It employs a hybrid TP-EP partitioner to optimize communication overhead and introduces a fused AR-A2A communication algorithm to enhance TTFT, ITL and throughput. MixServe's design is guided by theoretical analysis and practical considerations, ensuring efficient resource utilization and low latency. Our evaluation on mainstream MoE models such as DeepSeek-R1 and Qwen3 demonstrates that MixServe achieves significant performance improvements in MoE model serving, making it a valuable tool for deploying large-scale LLMs. We hope MixServe will contribute to the efficient deployment of MoE models in real-world applications.

% use the IEEEtran bibliography style
\bibliographystyle{IEEEtran}
\bibliography{ref}

\begin{thebibliography}{10}
\providecommand{\url}[1]{#1}
\csname url@samestyle\endcsname
\providecommand{\newblock}{\relax}
\providecommand{\bibinfo}[2]{#2}
\providecommand{\BIBentrySTDinterwordspacing}{\spaceskip=0pt\relax}
\providecommand{\BIBentryALTinterwordstretchfactor}{4}
\providecommand{\BIBentryALTinterwordspacing}{\spaceskip=\fontdimen2\font plus
\BIBentryALTinterwordstretchfactor\fontdimen3\font minus \fontdimen4\font\relax}
\providecommand{\BIBforeignlanguage}[2]{{%
\expandafter\ifx\csname l@#1\endcsname\relax
\typeout{** WARNING: IEEEtran.bst: No hyphenation pattern has been}%
\typeout{** loaded for the language `#1'. Using the pattern for}%
\typeout{** the default language instead.}%
\else
\language=\csname l@#1\endcsname
\fi
#2}}
\providecommand{\BIBdecl}{\relax}
\BIBdecl

\bibitem{DeepSeek-V3}
\BIBentryALTinterwordspacing
DeepSeek-AI \emph{et~al.}, ``Deepseek-v3 technical report,'' 2025. [Online]. Available: \url{https://arxiv.org/abs/2412.19437}
\BIBentrySTDinterwordspacing

\bibitem{DeepSeek-R1}
\BIBentryALTinterwordspacing
------, ``Deepseek-r1: Incentivizing reasoning capability in llms via reinforcement learning,'' 2025. [Online]. Available: \url{https://arxiv.org/abs/2501.12948}
\BIBentrySTDinterwordspacing

\bibitem{Qwen3}
\BIBentryALTinterwordspacing
A.~Yang \emph{et~al.}, ``Qwen3 technical report,'' 2025. [Online]. Available: \url{https://arxiv.org/abs/2505.09388}
\BIBentrySTDinterwordspacing

\bibitem{liu2025moeparallelfoldingheterogeneous}
\BIBentryALTinterwordspacing
D.~Liu \emph{et~al.}, ``Moe parallel folding: Heterogeneous parallelism mappings for efficient large-scale moe model training with megatron core,'' 2025. [Online]. Available: \url{https://arxiv.org/abs/2504.14960}
\BIBentrySTDinterwordspacing

\bibitem{Megatron-LM}
\BIBentryALTinterwordspacing
M.~Shoeybi \emph{et~al.}, ``Megatron-lm: Training multi-billion parameter language models using model parallelism,'' 2020. [Online]. Available: \url{https://arxiv.org/abs/1909.08053}
\BIBentrySTDinterwordspacing

\bibitem{Switch-Transformer}
\BIBentryALTinterwordspacing
W.~Fedus, B.~Zoph, and N.~Shazeer, ``Switch transformers: Scaling to trillion parameter models with simple and efficient sparsity,'' 2022. [Online]. Available: \url{https://arxiv.org/abs/2101.03961}
\BIBentrySTDinterwordspacing

\bibitem{vLLM}
\BIBentryALTinterwordspacing
W.~Kwon \emph{et~al.}, ``Efficient memory management for large language model serving with pagedattention,'' in \emph{Proceedings of the 29th Symposium on Operating Systems Principles}, ser. SOSP '23.\hskip 1em plus 0.5em minus 0.4em\relax New York, NY, USA: Association for Computing Machinery, 2023, p. 611–626. [Online]. Available: \url{https://doi.org/10.1145/3600006.3613165}
\BIBentrySTDinterwordspacing

\bibitem{Tutel}
\BIBentryALTinterwordspacing
C.~Hwang \emph{et~al.}, ``Tutel: Adaptive mixture-of-experts at scale,'' 2023. [Online]. Available: \url{https://arxiv.org/abs/2206.03382}
\BIBentrySTDinterwordspacing

\bibitem{ShareGPT-v3}
{OpenChat Team}, ``Openchat sharegpt v3,'' \url{https://huggingface.co/datasets/openchat/openchat_sharegpt_v3}, 2023, shareGPT dataset for training OpenChat V3 series. Licensed under MIT. Accessed: 2025-08-20.

\bibitem{GShard}
\BIBentryALTinterwordspacing
D.~Lepikhin \emph{et~al.}, ``Gshard: Scaling giant models with conditional computation and automatic sharding,'' 2020. [Online]. Available: \url{https://arxiv.org/abs/2006.16668}
\BIBentrySTDinterwordspacing

\bibitem{DeepSpeed-MoE}
\BIBentryALTinterwordspacing
S.~Rajbhandari \emph{et~al.}, ``Deepspeed-moe: Advancing mixture-of-experts inference and training to power next-generation ai scale,'' 2022. [Online]. Available: \url{https://arxiv.org/abs/2201.05596}
\BIBentrySTDinterwordspacing

\bibitem{SmartMoE}
\BIBentryALTinterwordspacing
M.~Zhai \emph{et~al.}, ``{SmartMoE}: Efficiently training {Sparsely-Activated} models through combining offline and online parallelization,'' in \emph{2023 USENIX Annual Technical Conference (USENIX ATC 23)}.\hskip 1em plus 0.5em minus 0.4em\relax Boston, MA: USENIX Association, Jul. 2023, pp. 961--975. [Online]. Available: \url{https://www.usenix.org/conference/atc23/presentation/zhai}
\BIBentrySTDinterwordspacing

\bibitem{Lina}
\BIBentryALTinterwordspacing
J.~Li \emph{et~al.}, ``Accelerating distributed {MoE} training and inference with lina,'' in \emph{2023 USENIX Annual Technical Conference (USENIX ATC 23)}.\hskip 1em plus 0.5em minus 0.4em\relax Boston, MA: USENIX Association, Jul. 2023, pp. 945--959. [Online]. Available: \url{https://www.usenix.org/conference/atc23/presentation/li-jiamin}
\BIBentrySTDinterwordspacing

\bibitem{Megatron-Core}
{NVIDIA Corporation}, ``Megatron-lm: Ongoing research training transformer models at scale,'' \url{https://github.com/NVIDIA/Megatron-LM}, 2024, accessed: 2025-08-20.

\bibitem{Orca}
\BIBentryALTinterwordspacing
G.-I. Yu \emph{et~al.}, ``Orca: A distributed serving system for {Transformer-Based} generative models,'' in \emph{16th USENIX Symposium on Operating Systems Design and Implementation (OSDI 22)}.\hskip 1em plus 0.5em minus 0.4em\relax Carlsbad, CA: USENIX Association, Jul. 2022, pp. 521--538. [Online]. Available: \url{https://www.usenix.org/conference/osdi22/presentation/yu}
\BIBentrySTDinterwordspacing

\bibitem{Llumnix}
\BIBentryALTinterwordspacing
B.~Sun \emph{et~al.}, ``Llumnix: Dynamic scheduling for large language model serving,'' in \emph{18th USENIX Symposium on Operating Systems Design and Implementation (OSDI 24)}.\hskip 1em plus 0.5em minus 0.4em\relax Santa Clara, CA: USENIX Association, Jul. 2024, pp. 173--191. [Online]. Available: \url{https://www.usenix.org/conference/osdi24/presentation/sun-biao}
\BIBentrySTDinterwordspacing

\bibitem{Alpa}
\BIBentryALTinterwordspacing
L.~Zheng \emph{et~al.}, ``Alpa: Automating inter- and {Intra-Operator} parallelism for distributed deep learning,'' in \emph{16th USENIX Symposium on Operating Systems Design and Implementation (OSDI 22)}.\hskip 1em plus 0.5em minus 0.4em\relax Carlsbad, CA: USENIX Association, Jul. 2022, pp. 559--578. [Online]. Available: \url{https://www.usenix.org/conference/osdi22/presentation/zheng-lianmin}
\BIBentrySTDinterwordspacing

\bibitem{AlpaServe}
\BIBentryALTinterwordspacing
Z.~Li \emph{et~al.}, ``{AlpaServe}: Statistical multiplexing with model parallelism for deep learning serving,'' in \emph{17th USENIX Symposium on Operating Systems Design and Implementation (OSDI 23)}.\hskip 1em plus 0.5em minus 0.4em\relax Boston, MA: USENIX Association, Jul. 2023, pp. 663--679. [Online]. Available: \url{https://www.usenix.org/conference/osdi23/presentation/li-zhouhan}
\BIBentrySTDinterwordspacing

\bibitem{DistServe}
\BIBentryALTinterwordspacing
Y.~Zhong \emph{et~al.}, ``{DistServe}: Disaggregating prefill and decoding for goodput-optimized large language model serving,'' in \emph{18th USENIX Symposium on Operating Systems Design and Implementation (OSDI 24)}.\hskip 1em plus 0.5em minus 0.4em\relax Santa Clara, CA: USENIX Association, Jul. 2024, pp. 193--210. [Online]. Available: \url{https://www.usenix.org/conference/osdi24/presentation/zhong-yinmin}
\BIBentrySTDinterwordspacing

\bibitem{Sarathi-Serve}
\BIBentryALTinterwordspacing
A.~Agrawal \emph{et~al.}, ``Taming {Throughput-Latency} tradeoff in {LLM} inference with {Sarathi-Serve},'' in \emph{18th USENIX Symposium on Operating Systems Design and Implementation (OSDI 24)}.\hskip 1em plus 0.5em minus 0.4em\relax Santa Clara, CA: USENIX Association, Jul. 2024, pp. 117--134. [Online]. Available: \url{https://www.usenix.org/conference/osdi24/presentation/agrawal}
\BIBentrySTDinterwordspacing

\bibitem{MegaScale-Infer}
\BIBentryALTinterwordspacing
R.~Zhu \emph{et~al.}, ``Megascale-infer: Serving mixture-of-experts at scale with disaggregated expert parallelism,'' 2025. [Online]. Available: \url{https://arxiv.org/abs/2504.02263}
\BIBentrySTDinterwordspacing

\end{thebibliography}

\end{document}